# Hydrogen embrittlement of twinning-induced plasticity steels: contribution of segregation to twin boundaries

Heena Khanchandani[1], Rolf Rolli[2], Hans-Christian Schneider[2], Christoph Kirchlechner[2], Baptiste Gault[1,3*]

[1]Max-Planck-Institut für Eisenforschung; Düsseldorf; Germany.
[2]Institute for Applied Materials, Karlsruher Institute of Technology; Karlsruhe; Germany.
[3]Imperial College; London; United Kingdom.
*Corresponding author: b.gault@mpie.de

**Abstract:** Metallic materials, especially steel, underpin transportation technologies. High-manganese twinning induced plasticity (TWIP) austenitic steels exhibit exceptional strength and ductility from twins, low-energy microstructural defects that form during plastic loading. Their high-strength could help light-weighting vehicles, and hence cut carbon emissions. TWIP steels are however very sensitive to hydrogen embrittlement that causes dramatic losses of ductility and toughness leading to catastrophic failure of engineering parts. Here, we examine the atomic-scale chemistry and interaction of hydrogen with twin boundaries in a model TWIP steel by using isotope-labelled atom probe tomography, using tritium to avoid overlap with residual hydrogen. We reveal co-segregation of tritium and, unexpectedly, oxygen to coherent twin boundaries, and discuss their combined role in the embrittlement of these promising steels.

Candidate metallic alloys allowing new design of light-weight vehicles in support of the *net-zero-carbon* transition must exhibit novel combinations of high-strength and ductility. High-manganese (above 20 wt.%), face centered cubic (FCC), twinning induced plasticity (TWIP) austenitic steels [1] are a potential scalable material solution. Upon loading, TWIP steels deform through dislocation glide along with the continuous formation of mechanical twins [2]. A twin is a crystal related to its parent by a symmetry operation. Twins are often seen as a local rotation of the lattice about one of its main symmetry axes, and often result in a change in the sequence in the stacking of the atoms. Twin boundaries are low energy structural defects that are found in e.g. minerals [3,4] or metals [2,5,6]. In TWIP steels, through the 'dynamic Hall-Petch effect' [7], twin boundaries hinder dislocation motion and provide tensile strength up to 800 MPa while maintaining a high ductility such that the elongation to failure is up to 100% [8].

A key obstacle to the widespread application of TWIP steel is their high susceptibility to hydrogen embrittlement (HE) [9–12]. Hydrogen is the lightest, most abundant and most mobile element [10,11] and is projected to be an energy vector for contribution to replacing fossil fuels in transportation or electricity generation [13], yet its deleterious influence on the mechanical properties of materials has been known for over a century [14]. HE leads to a drop in e.g. toughness [15,16] that has led to numerous catastrophic failures [11,16] such as the oil spillage in the Gulf of Mexico in 2012 [17] and the collapse of the San Francisco – Oakland Bay Bridge in 2013 [18].

Exposure of TWIP steel to hydrogen under loading conditions also leads to premature failure [9]. Hydrogen ingress is associated with an increase in twin density in individual twin bundles [19] and a reduction in twin thickness that leads to their embrittlement [20]. Overcoming HE requires understanding the interaction of hydrogen with grain and twin boundaries in TWIP steels at the atomic scale in order to determine their role in HE [9].

Here, we quantified the solute segregation at coherent and incoherent annealing twin boundaries [8] in a recrystallized, model Fe 28Mn 0.3C (wt.%) TWIP steel by using atom probe tomography (APT) correlated with electron backscatter diffraction (EBSD) and electron channeling contrast imaging (ECCI). We selected annealing twins as a proxy as nanoscale analytical imaging of deformation twins is much more challenging. We also introduce isotopic-labelling with tritium ($^3$H) to facilitate quantification by APT [21]. To circumvent the overlap of hydrogen from residual gases [22–25], samples are often charged with deuterium, yet the detection of $H_2^+$ still hinders H-quantification [25,26].

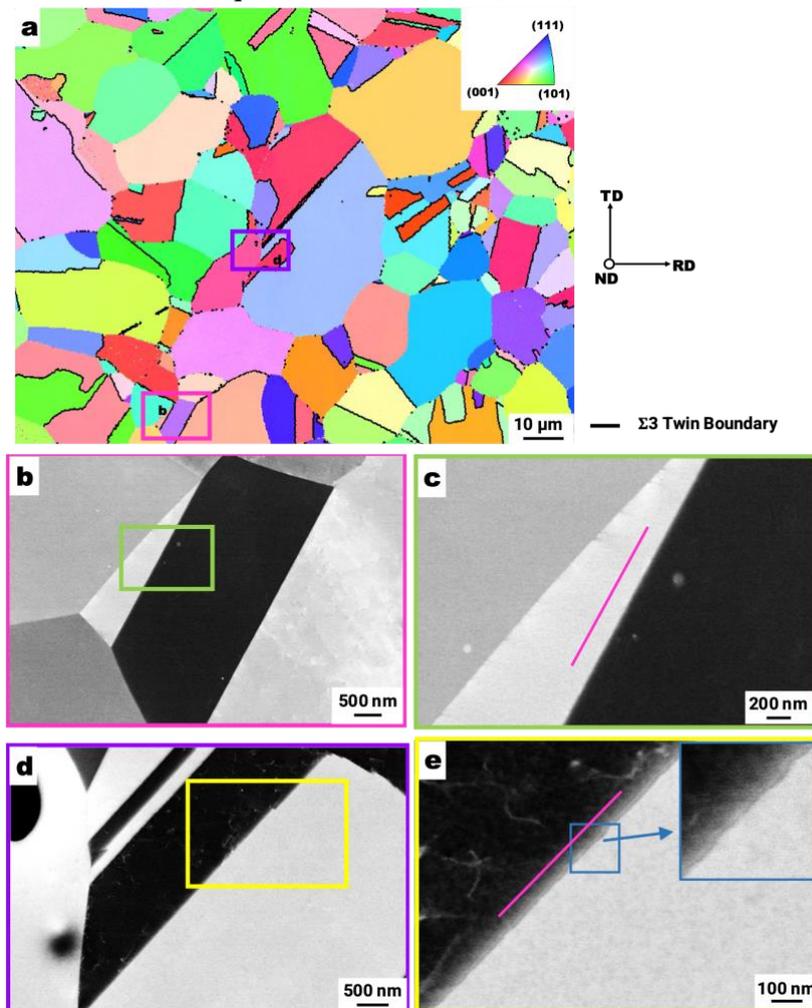

*Fig. 1.* *(a) Electron backscatter diffraction-inverse pole figure (EBSD-IPF) map highlighting the Σ3 twin boundaries in black. (b) The electron channeling contrast (ECC) image of a coherent Σ3 twin boundary highlighted by the pink box in (a), whose image at a higher magnification is shown in (c) confirming its straight structure. (d) The ECC image of an incoherent Σ3 twin boundary highlighted by the purple box in (a), with (e) showing its image at higher magnification in which the "hills and valleys" morphology of the boundary is illustrated in the inset.*

The model Fe 28Mn 0.3C (wt.%) TWIP steel was produced by strip casting [27] and homogenized at 1150 °C for 2 hours [28]. It was cold-rolled to achieve a 50% thickness reduction, and subjected to a recrystallization annealing treatment at 800 °C for 20 minutes, followed by water cooling to room temperature. EBSD was performed using a Zeiss Sigma 500 SEM equipped with an EDAX/TSL system with a Hikari camera at an accelerating voltage of 15 kV, a beam current of 9 nA, a scan step size of 0.5 µm, a specimen tilt angle of 70°, and a working distance of 14 mm [29]. The ECCI was performed using a Zeiss Merlin SEM, equipped with a retractable 4Q backscattered electrons detector at an accelerating voltage of 30 kV, a beam current of 2 nA and a working distance of 7 mm [30]. APT specimens were prepared by site-specific lift-out procedure [31] on a FEI Helios NanoLab 600i dual-beam FIB/SEM equipped with a Hikari camera for performing transmission Kikuchi diffraction (TKD) with a step size of 20 nm. EBSD and TKD data analyses were performed using OIM Data Analysis 7.0.1 (EDAX Inc.) software. APT experiments were conducted on either a LEAP 5000 XS or XR instrument (CAMECA Instruments Inc. Madison, WI, USA), in voltage pulsing mode at a temperature of 70 K, 15-20 % pulse fraction, 200 kHz pulse repetition rate and 0.5 % detection rate.

Coherent and incoherent twin boundaries differ in their habit plane and hence the grain boundary energy [32,33]. Details of their morphology, such as nanoscale steps [32,34], cannot be readily seen from a low-resolution EBSD inverse pole figure (IPF) map, a coherent boundary must be carefully identified through trace analysis in the EBSD. The structure of several Σ3 twin boundaries was hence investigated by ECCI and a coherent and an incoherent Σ3 twin boundary are studied. Fig. 1a shows the EBSD-IPF map with black color-coded twin boundaries according to Brandon's criterion [35]. Twins represent 36.6% of all grain boundaries. The ECC image of a coherent twin boundary is shown in Fig. 1b, with Fig. 1c a close-up at higher magnification showcasing a straight structure. The (111) pole figures from the abutting grains and the trace analysis are reported in supplementary Fig. S1a. The incoherent twin boundary, Fig. 1d exhibits a "hills and valleys" morphology [36] at the nanoscale, Fig. 1e, previously reported for incoherent twin boundaries [37,38]. Supplementary Fig. S1b shows the corresponding (111) pole figures and trace analysis [32].

Since HE and corrosion mechanisms are interrelated [39], we immersed the sample into deionized water for 5 minutes. Supplementary Fig. S2a shows the EBSD-IPF map after this corrosion test, with black color-coded twin boundaries. Coherent twins are low energy boundaries [32,33], and hence not expected to be reactive, yet the yellow box in supplementary Fig. S2b indicates a corroded coherent twin boundary, whose coherency was verified by the pole figure trace analysis in supplementary Fig. S3, suggesting a high embrittlement sensitivity of coherent twins. In addition, a bulk tensile-test sample was electrochemically loaded with hydrogen and tested to fracture [40], and the micrographs in supplementary Fig. S4 give some indications of crack propagation in the vicinity or through twins, i.e. making them microstructural locations of interest to understand hydrogen embrittlement.

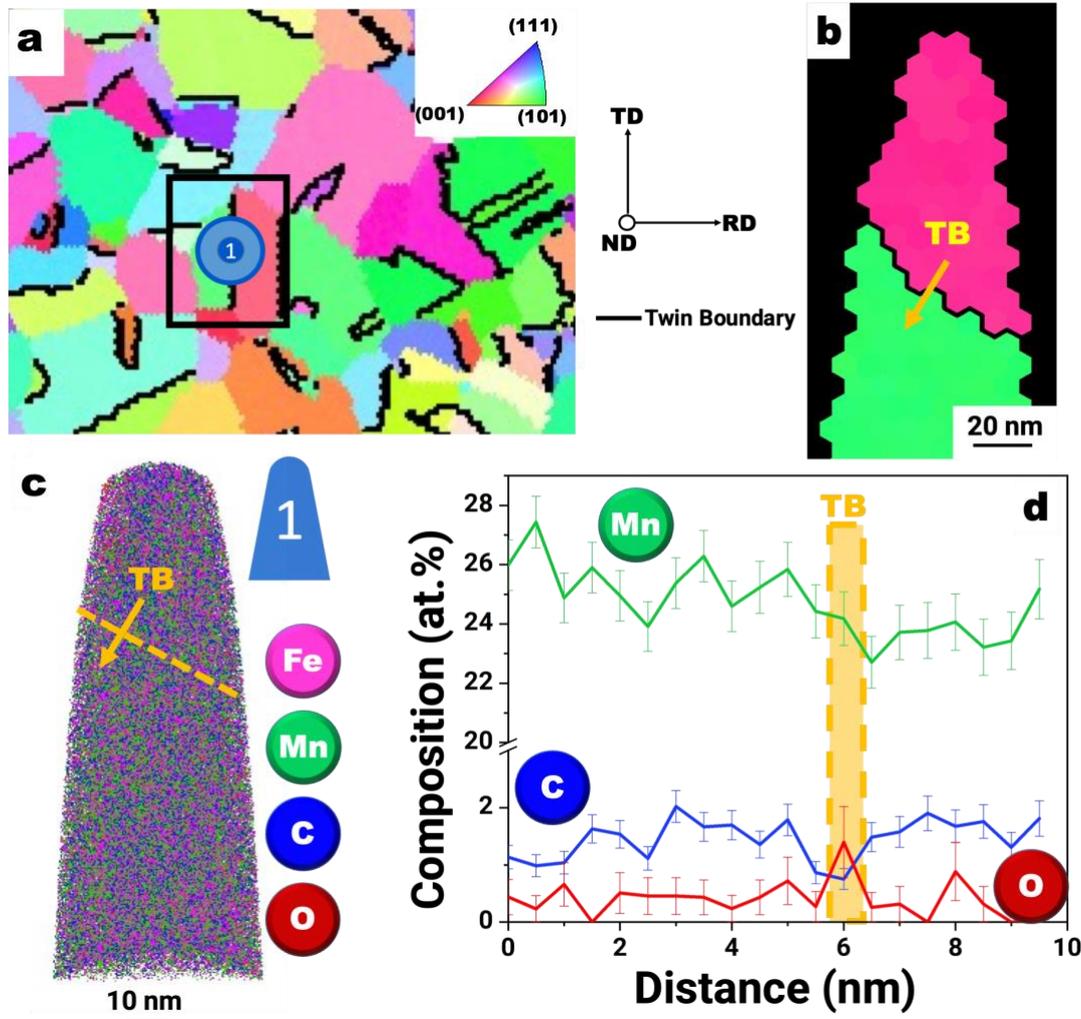

*Fig. 2.* *(a) Electron backscatter diffraction-inverse pole figure (IPF) map with respect to normal direction (ND) out of the plane, a black box highlightsing the investigated boundary in pink black box. (b) Transmission Kikuchi diffraction-IPF map indicating the twin boundary (TB). (c) 3-D elemental map showing iron (Fe), manganese (Mn), carbon (C) and oxygen (O) atoms following the APT reconstruction. (d) Composition profile calculated with 0.5 nm bin width at the twin boundary highlighting the oxygen enrichment and manganese depletion.*

The coherent Σ3 twin boundary, supplementary Fig. S5a, is highlighted by a black box in the EBSD-IPF map in Fig. 2a. It was selected for correlative investigation by transmission Kikuchi diffraction (Fig. 2b) and APT. The 3-D elemental map is shown in Fig. 2c. The local point density fluctuation shown in supplementary Fig. S5b, allowed us to quantify precisely the composition across the boundary. Fig. 2d evidences no segregation of carbon at this coherent Σ3 twin boundary, which can be ascribed to its low-energy, straight and flat structure. In contrast, there is a slight depletion of manganese (approx. $-3 \pm 0.8$ at.%) which can be expected from the Mn increasing trend in the stacking fault energy associated to Mn above approx. 10 wt.% reported previously [41,42]. We can also report an unexpected strong oxygen enrichment of $1.4 \pm 0.6$ at.%.

Subsequently, tritium charging was performed on well-polished samples of dimensions 2×2×1 mm$^3$ by using a gas mixture of 500 ppm of tritium in hydrogen. The tritium charging was performed at 70 °C at 4 bar pressure for 6 hours. Thermal desorption analysis, Fig. S6, was performed using a DDH 8 type tritium flow proportional detector equipped with an MKS RGA Microvision 2 quadrupole mass spectrometer. The sample was heated at a heating rate of 1 °C/min from room temperature up-to 300 °C and the corresponding tritium desorption rate was measured in Bq/s and re-calculated in wt. ppm/s. This shows a tritium content of $1.5 \times 10^{-2}$ wt. ppm.

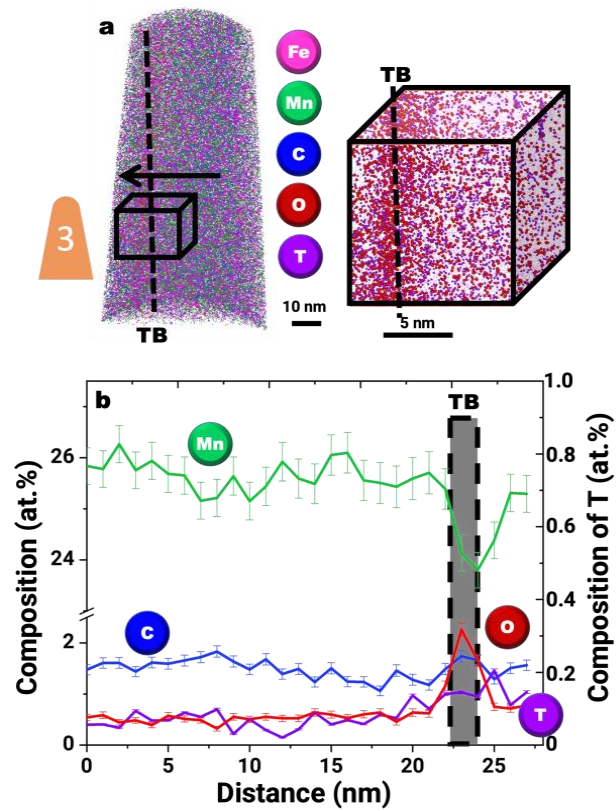

*Fig. 3.* *(a) 3-D elemental map showing iron (Fe), manganese (Mn), carbon (C), oxygen (O) and tritium (T) atoms following the APT reconstruction. (b) Composition profile calculated with 1 nm bin width at the tritium-charged twin boundary (TB) highlighting the oxygen and tritium enrichment. Manganese depletion suggests that it is a coherent Σ3 twin boundary.*

The APT reconstruction of a tritium-charged twin boundary in Fig. 3a and the composition profile across the boundary in Fig. 3b evidences a similar oxygen enrichment of $2.3 \pm 0.1$ at.% and a tritium segregation of $0.15 \pm 0.03$ at.%. In comparison with the uncharged material, Fig. 2, the manganese depletion (approx. $-3 \pm 0.13$ at.%) allows us to infer that this is a coherent Σ3 twin boundary – the analysis of incoherent twin boundaries, as showcased in supplementary Fig. S7 shows a different segregation behavior, with no oxygen but carbon segregation. No detectable oxygen levels are found at incoherent twins or in the bulk of the grains. The oxygen can originate either from residuals in the alloy, the bulk chemical analysis measured 0.026 wt.% of oxygen, or from air-exposure between preparation and analysis. Oxygen segregation behavior to a coherent boundary is difficult to rationalize, in particular in contrast to carbon's. It may be due to a competitive occupation of most favorable trapping sites at grain boundaries by carbon leaving only the coherent twins for oxygen to segregate, or is related to the Mn depletion. This will grant further investigations by e.g. atomistic simulations.

The relatively slower diffusivity of tritium compared to hydrogen [43] helped maintain sufficient tritium inside the sample to be detectable by APT, despite four weeks between charging and analysis. Although complex cryogenic transfer workflows [26,44] were proposed for analyzing hydrogen by APT, tritium segregation was revealed here without, a possibility discussed in Ref. [25]. The reliability of hydrogen detection by APT strongly depends on experimental conditions [45], our analyses in supplementary Fig. S8 support that the tritium content at the coherent twin boundary is related to the charging and not arising from residual gases. Koyama and coworkers reported hydrogen at twin boundaries in a TWIP steel by scanning Kelvin probe force microscopy [46], which is in agreement with our observations, although the coherency of the studied twin boundaries was not reported. It has been proposed that hydrogen reduces the stacking fault energy in austenitic steels [47,48] which causes a high twinability at the grain boundary leading to the formation of more deformation twins, and their exposure to hydrogen can be subject to decohesion and facilitate crack propagation. The hydrogen embrittlement hence occurs by hydrogen-enhanced twin boundary decohesion. First-principles studies also propose that the hydrogen at Σ3 twins leads to a mild increase in the grain boundary energy [49], facilitating grain boundary decohesion and hence the critical stress to fracture the material, similar to what was previously observed for e.g. Al-alloys [50].

To summarize, tritiation revealed that the twin boundaries that underpin the extraordinarily high tensile strength and ductility of TWIP steels can trap hydrogen and hence enhance their susceptibility to HE. However, our results also suggest an unforeseen sensitivity of twins to oxygen-containing environments, including water, Fig. S2. Recent work on stress-corrosion cracking of Ti or Al alloys had already indicated some unexpected ingress and role of oxygen [51,52]. A higher level of residual O in austenitic steels was previously linked to a higher tendency for void formation and an associated loss of toughness [53] which could further contribute to the embrittlement. *Ab initio* calculations suggest that hydrogen binds very strongly to oxygen in austenite [54], and we propose this may here though that oxygen also contributes to increasing the level of hydrogen at the twins, thereby possibly facilitating causing the embrittlement of TWIP steels. Our new insights will now motivate the detailed study of the combined effect of O and H on the cohesion of twins.

**Acknowledgments**

The authors would like to thank Uwe Tezins, Andreas Strum and Christian Bross for their support to the FIB and APT facilities at MPIE. Ms. Monika Nellessen and Ms. Katja Angenendt are gratefully acknowledged for their support to the metallography sample preparation lab and SEM facilities at MPIE. We thank Prof. Dierk Raabe, Drs. Stefan Zaefferer and Dirk Ponge for helpful discussions. H.K. and B.G. acknowledge the financial support from the ERC-CoG-SHINE-771602.

**Supplementary Materials**

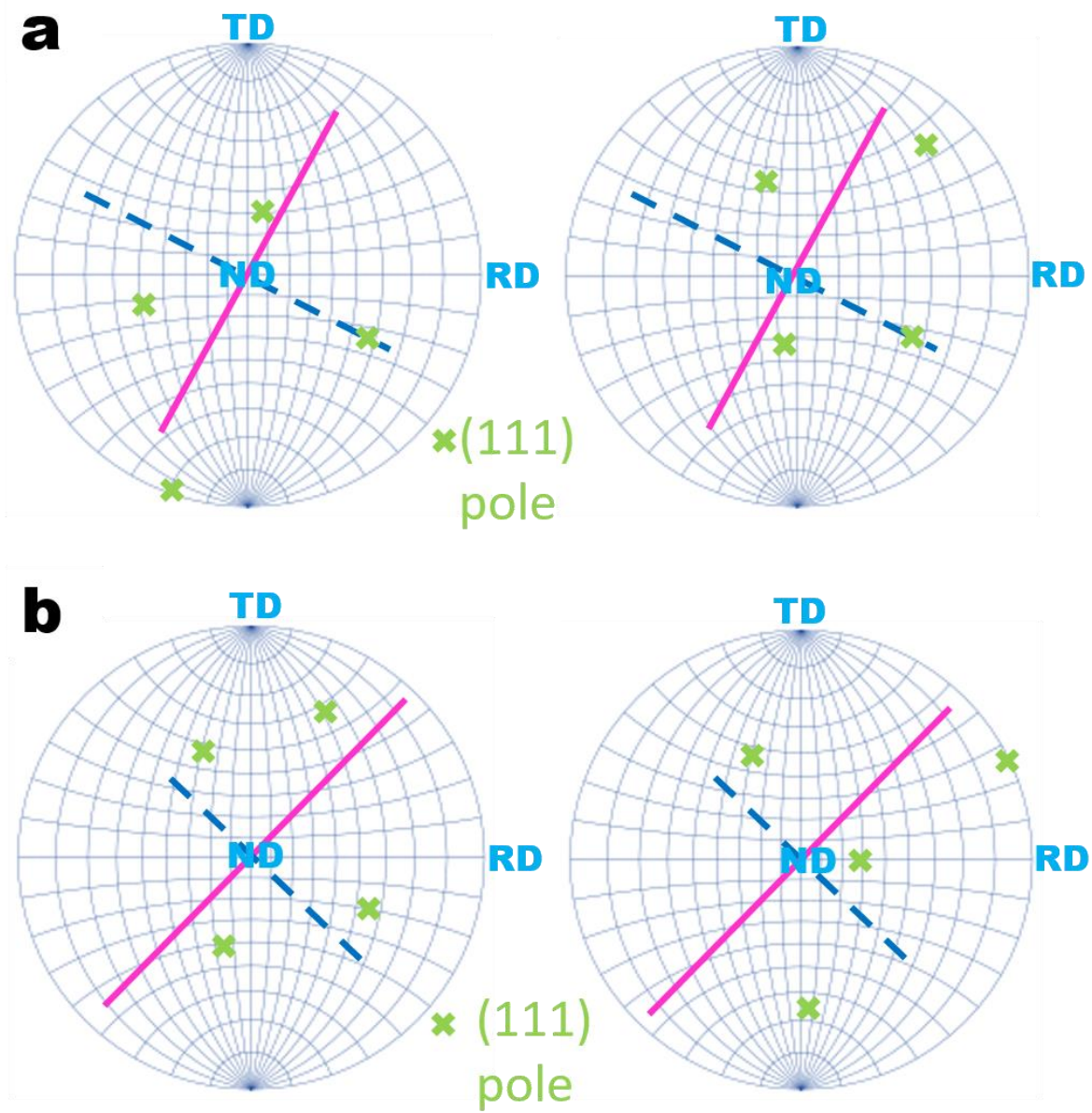

**Supplementary fig. S1. (111) pole figures from the abutting grains for determining the coherency of Σ3 twin boundaries.** The twin boundary trace is indicated in pink, while the boundary normal vector is highlighted by the dotted blue line. (111) poles are indicated with green dots. (a) The boundary normal vector passes through the (111) pole common to both abutting grains, thereby verifying the coherency of the boundary [31] shown in Fig. 1b. (b) The boundary normal vector does not pass through the common (111) pole, indicating that the boundary displayed in Fig. 1d is an incoherent Σ3 twin boundary [31,52].

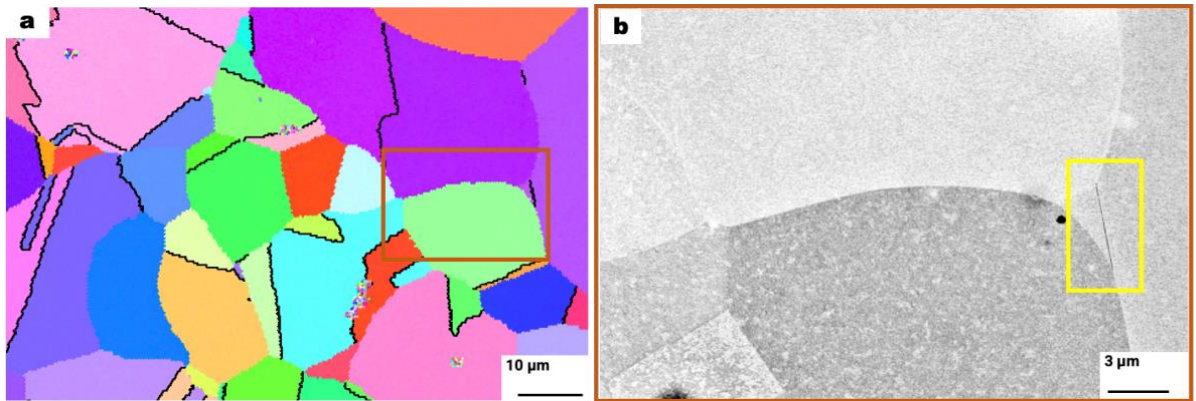

**Supplementary fig. S2. Water-based corrosion of coherent twins.** (a) EBSD-IPF map of the studied TWIP steel sample after performing the corrosion test highlighting the Σ3 twin boundaries in black. (b) A coherent Σ3 twin boundary indicated in yellow which is corroded. All EBSD-IPF maps are with respect to the normal direction out of the plane. ND – normal direction, TD – transverse direction, RD – rolling direction.

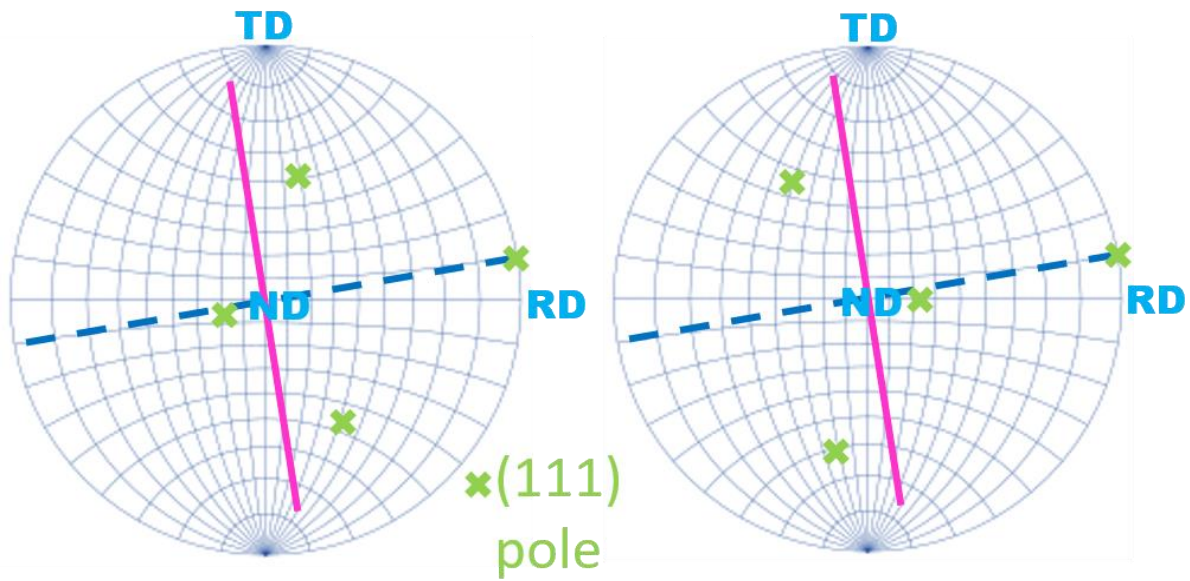

**Supplementary fig. S3. (111) pole figures corresponding to the coherent Σ3 twin boundary shown in supplementary Fig. S2b verifying its coherency.**

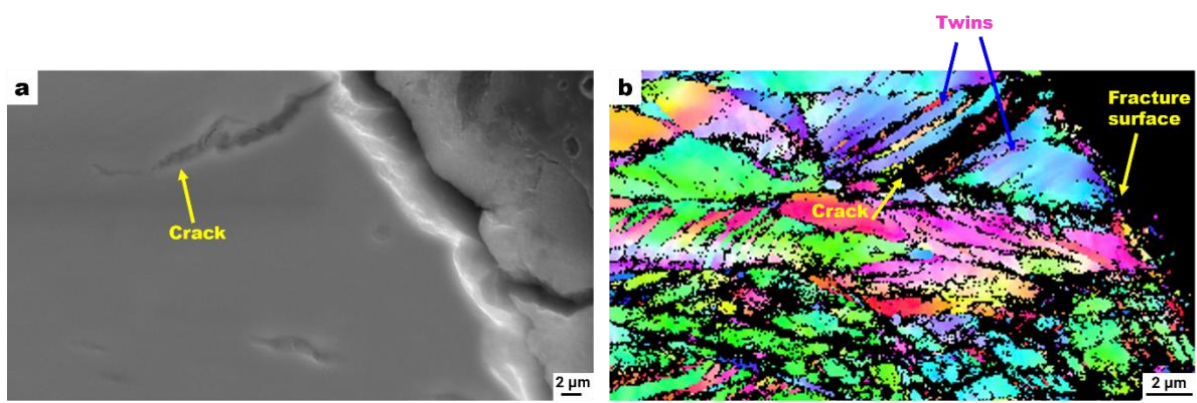

**Supplementary fig. S4. Fracture surface of the hydrogen-charged tensile specimen deformed to fracture.** (a) SEM image indicating a crack near the fracture surface. (b) Electron backscatter diffraction-inverse pole figure map with respect to normal direction (ND) out of the plane depicting twins near the crack.

The tensile specimen of the studied recrystallized TWIP steel sample was charged with hydrogen electrolytically in an aqueous solution of 0.05 M $H_2SO_4$, with 1.4 g/l of thiourea ($CH_4N_2S$) as hydrogen recombination binder for 5 days. After 5 days of charging, it was deformed to tensile fracture. The SEM image close to the fracture surface, Fig. S4a exhibited a crack which was mapped by EBSD whose IPF is shown in supplementary Fig. S4b. Twins were observed near the fractured surface. Approx. 37% of total grain boundaries are twins in the recrystallized material and their presence near the fracture surface in Fig. S4b suggests the decohesion of twins on the introduction of hydrogen into the material.

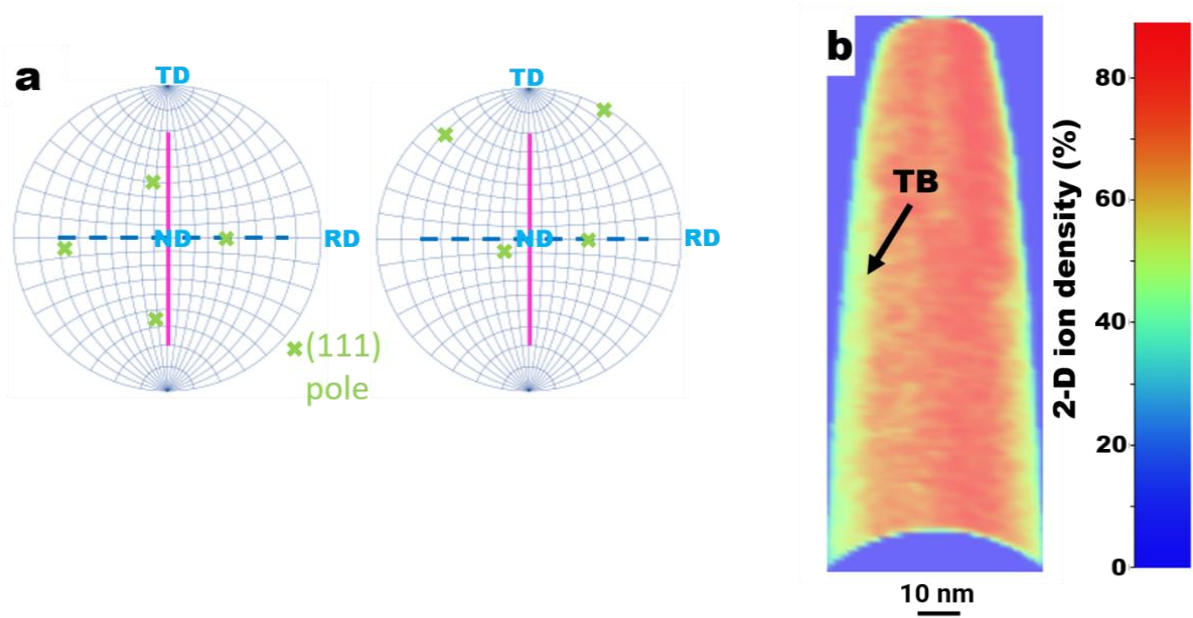

**Supplementary fig. S5. Coherent Σ3 twin boundary analyzed in Fig. 2.** (a) (111) pole figures from the abutting grains verifying its coherency. (b) 2-D ion density (ion.%) composition plot indicating the studied twin boundary (TB).

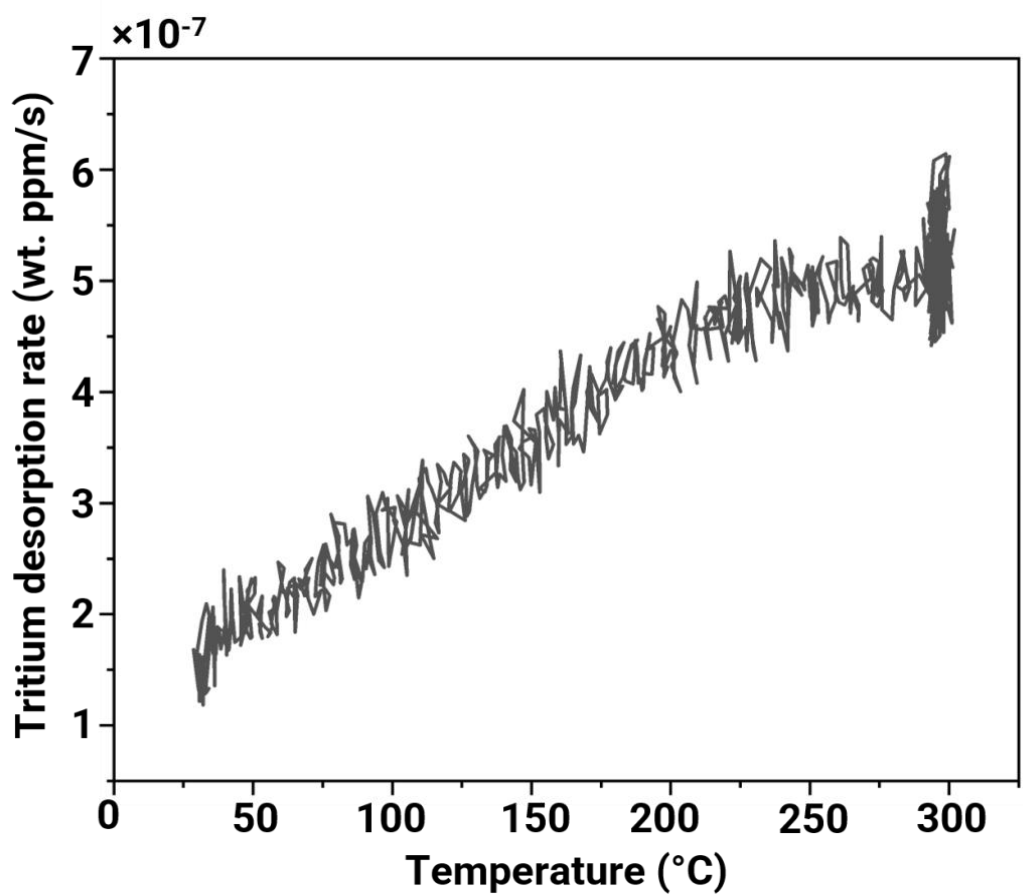

**Supplementary fig. S6. Thermal desorption analysis curve of the tritium-charged sample.**
Tritium desorption rate calculated to wt. ppm/s is shown as a function of temperature (°C) from 30°C to 300°C.

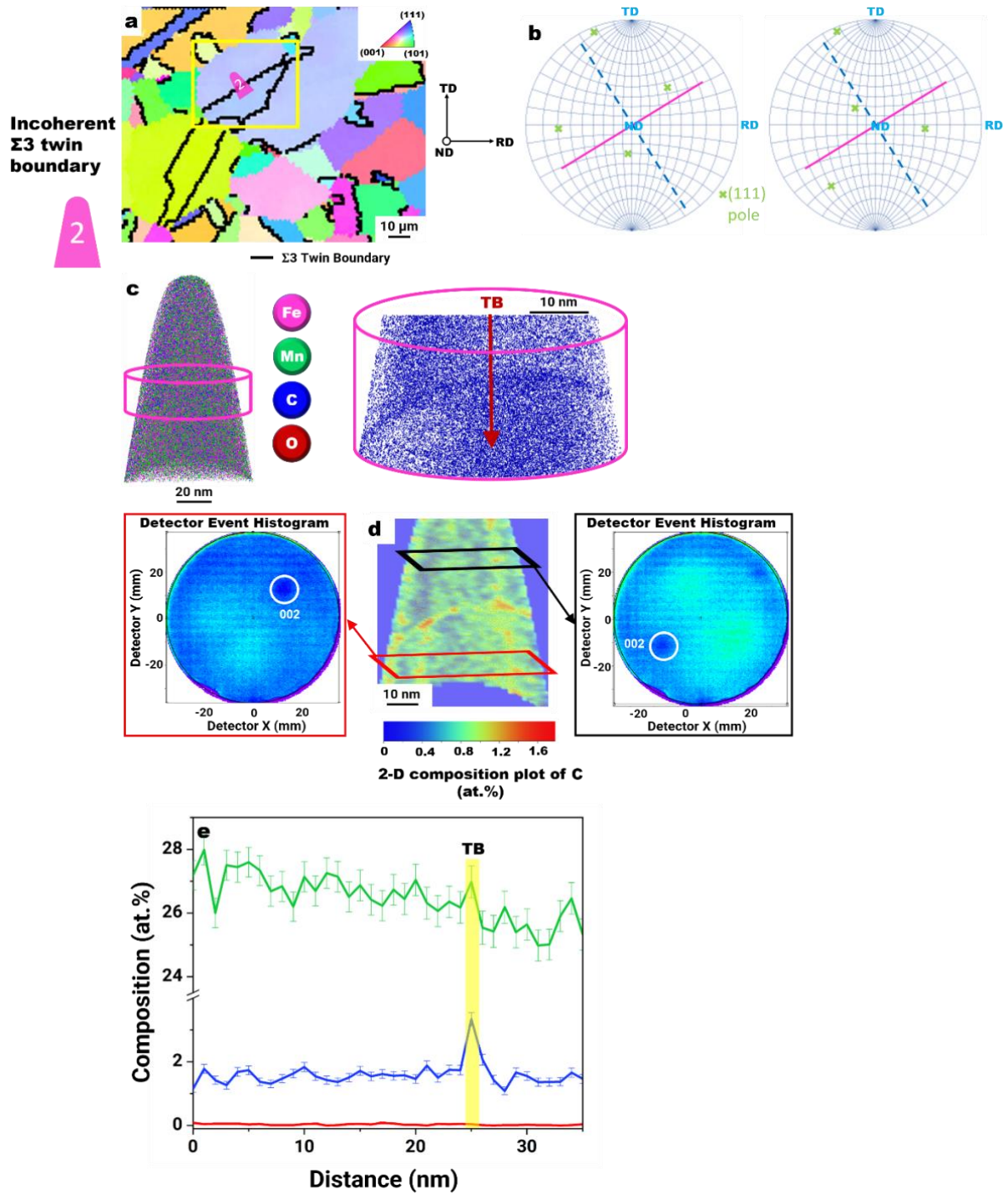

**Supplementary fig. S7. Incoherent Σ3 twin boundary analyzed by APT.** (a) Electron backscatter diffraction-inverse pole figure map with respect to normal direction (ND) out of the plane, highlighting the investigated boundary in yellow box. (b) (111) pole figures from the abutting grains verifying its incoherency. (c) 3-D elemental map showing iron (Fe), manganese (Mn), carbon (C) and oxygen (O) atoms following the APT reconstruction. (d) 2-D carbon composition map (at.%) with detector event maps showing the inversion of (002) pole indexed from the abutting grains which also strongly suggests that it is a Σ3 twin boundary [53]. (e) The composition profile with 1 nm bin width across the incoherent Σ3 twin boundary (TB) highlighting the carbon enrichment of 3.3 ± 0.2 at.%, with no manganese or oxygen enrichment. APT analysis of twin boundaries in TWIP steels showed no elemental segregation, but their coherency was not reported [54]. The faceting also strongly influences the local

chemistry of a grain boundary [33,55]. The ECC image of the incoherent Σ3 twin boundary at a higher magnification (Fig. 1e) indicates that the morphology of the boundary consisted of variations at the nanoscale, which explains the observed carbon segregation.

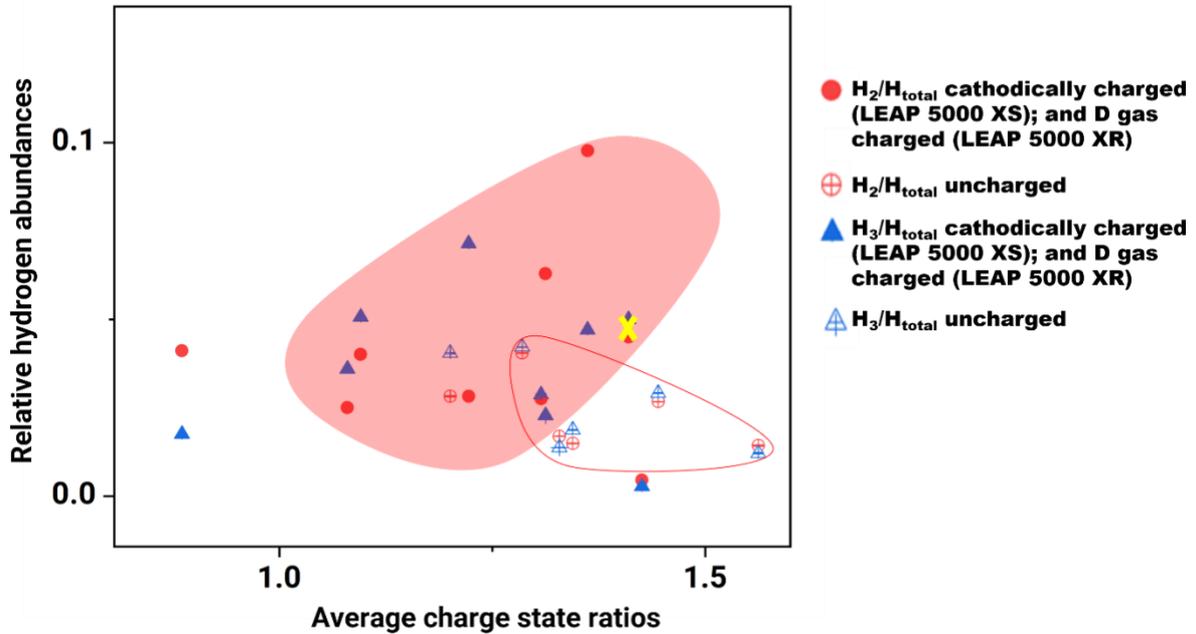

**Supplementary fig. S8. The position of tritium-charged coherent Σ3 twin boundary highlighted by a yellow 'X' in the master curve drawn in ref** [26]**.** We plotted a master curve of the average charge state ratios vs. relative hydrogen abundances from multiple datasets of the hydrogen-charged and the uncharged specimens in ref. [26]. The region highlighted in red contains data mostly from the hydrogen-charged specimens, as opposed to the region delineated by red which comprises of data mostly from the uncharged specimens. The relative $H_2$ and $H_3$ contents from the tritium-charged dataset are shown in the same master curve here, in which the position of tritium-charged coherent Σ3 twin boundary lies in the region with higher relative $H_3$ contents, as highlighted by a yellow *'X'*, which also confirms the tritium charging and a substantial tritium ingress into the studied sample. $H_3$ is detected at 3Da in an APT dataset, whose content is negligible in uncharged specimens while it could be H+D in the D gas charged specimens. The peak at 3Da was indexed as tritium in the tritium-charged specimen.

**Spatial distribution map analysis of the tritium-charged coherent Σ3 twin boundary**

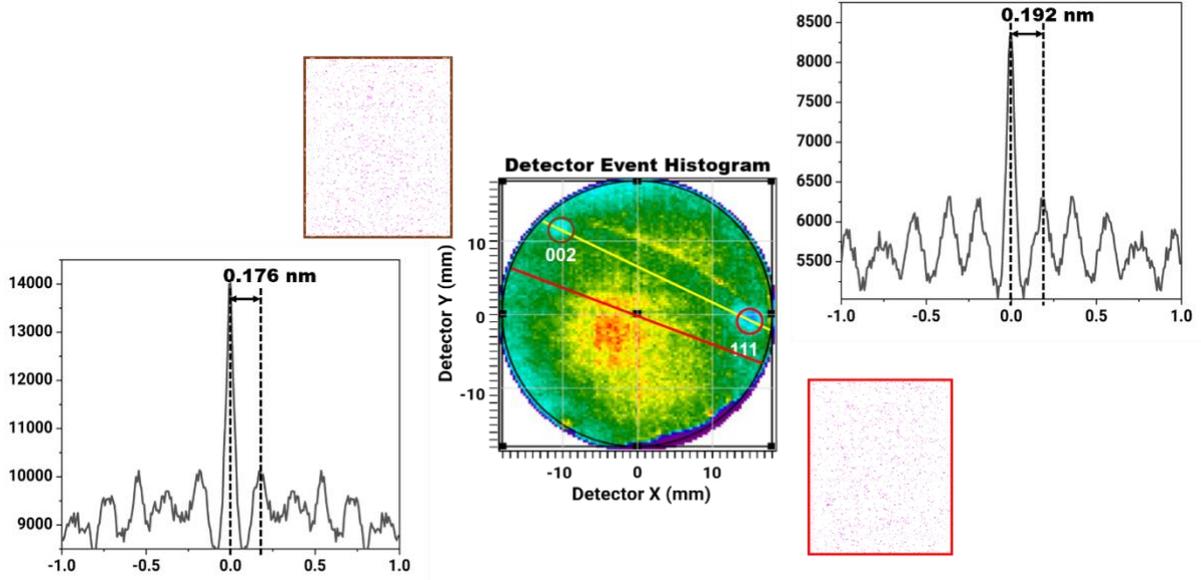

**Supplementary fig. S9. The detector event map depicting planes and spatial distribution maps of (002) and (111) poles of the tritium-charged coherent Σ3 twin boundary.**

The detector event map shown in supplementary Fig. S9 also confirms the presence of a twin boundary, while showing the planes and spatial distribution maps of (002) and (111) poles. The *d* spacing of the (002) plane obtained from its spatial distribution map was 0.176 nm (theoretical value – 0.18 nm) and that of the (111) plane was 0.192 nm (theoretical value – 0.207 nm).

The distance between crystallographic poles is proportional to the crystallographic angle between these directions [53,56]. The diameter of detector event map is 8.5 cm, indicated by a red line in supplementary Fig. S9 which corresponds to an angle of 60°. The distance between the two poles is 7.7 cm indicated by a yellow line. The angle between the two planes (111) and (002) can hence be calculated as 54.35°. Theoretically, the angle $\phi$ between two planes, ($h_1 k_1 l_1$) and ($h_2 k_2 l_2$) is given by:

$$cos\,\phi = \frac{h_1 h_2 + k_1 k_2 + l_1 l_2}{\sqrt{h_1^2 + k_1^2 + l_1^2}\sqrt{h_2^2 + k_2^2 + l_2^2}}$$

$$cos\,\phi = \frac{(1 \times 0) + (1 \times 0) + (1 \times 2)}{\sqrt{1+1+1}\sqrt{4+0+0}}$$

$$cos\,\phi = \frac{1}{\sqrt{3}}$$

$\phi = 54.73°$ which is consistent with the value calculated from the detector event map.